\documentclass[journal]{IEEEtran}
\pdfoutput=1
\usepackage{epstopdf}
\usepackage{amsmath,amssymb}
\usepackage{multicol}
\usepackage{graphicx,graphics,color,psfrag}
\usepackage{cite,balance}
\usepackage{algorithm}
\usepackage{accents}
\usepackage{caption}
\usepackage{bm}
\usepackage{url}
\usepackage{algorithmic}
\usepackage[english]{babel}
\usepackage{multirow}
\usepackage{enumerate}
\usepackage{cases}
\usepackage{stfloats}
\usepackage{dsfont}
\usepackage{color,soul}
\usepackage{amsfonts}
\usepackage{tcolorbox}
\usepackage{amsmath}
\usepackage{float}
\usepackage{siunitx}
\usepackage{xcolor}

\usepackage{cite,graphicx,amsmath,amssymb}
\usepackage{fancyhdr}
\usepackage{hhline}
\usepackage{graphicx,graphics}
\usepackage{array,color}
\usepackage{amsmath}
\usepackage{stfloats}
\usepackage[flushleft]{threeparttable}
\usepackage{booktabs}
\usepackage{amsthm}

\newcounter{MYtempeqncnt}

\ifCLASSINFOpdf

\else

\fi

\begin{document}
\captionsetup[figure]{labelfont={default},labelformat={default},labelsep=period,name={Fig.}}
\title{Wireless Control over Edge Networks: Joint User Association and Communication-Computation Co-Design}

\author{Zhilin~Liu, Yiyang~Li, Huijun Xing, Ye Zhang, Jie~Xu, and Shuguang Cui}

\maketitle
\begin{abstract}
This paper studies a wireless networked control system with multiple base stations (BSs) cooperatively coordinating the wireless control of a number of subsystems each consisting of a plant, a sensor, and an actuator. In this system, each sensor first offloads the sensing data to its associated BS, which then employs mobile edge computing (MEC) to process the data and sends the command signals back to the actuator for remote control. We consider the time-division-multiple-access (TDMA) service protocol among different BSs to facilitate the cascaded communication and computation process, in which different BSs implement the uplink data collection and downlink command broadcasting over orthogonal time slots. We also employ the massive multiple-input multiple-output (MIMO) at BSs, based on which each BS serves its associated sensors or actuators over the same time-frequency resources via spatial multiplexing.
Under this setup, we jointly design the association between BSs and sensors/actuators as well as the joint communication and computation resource allocation, with the objective of minimizing the closed-loop control latency of the multiple subsystems while ensuring their control stability. The optimization takes into account the transmission uncertainty caused by both the hyper reliable and low-latency communications (HRLLC) and the inter-user interference , as well as the communication and computation resource constraints at distributed nodes. To solve the challenging non-convex joint optimization problem, we develop an efficient algorithm by employing the techniques of alternating optimization and successive convex approximation (SCA). Numerical results show that the proposed joint BS-sensor/actuator association and resource allocation design significantly outperforms other heuristic schemes and frequency-division-multiple-access (FDMA) counterpart.
\end{abstract}

\begin{IEEEkeywords}
Wireless networked control, mobile edge computing (MEC), hyper reliable and low-latency communications (HRLLC), user association, resource allocation.
\end{IEEEkeywords}

\section{Introduction}
Wireless networked control is a key usage scenario for sixth-generation (6G) networks \cite{add1}, enabling industrial automation by allowing remote controllers, such as base stations (BSs), to wirelessly collect sensing data and send commands, enhancing system flexibility, scalability, and cost efficiency \cite{con}. Typically, BSs collect sensor data, process it via edge servers, and transmit commands to actuators for control. Efficient coordination of communication, computation, and control between BS controllers and subsystems is essential.

Ensuring stability and efficiency in wireless networked control requires low-latency, high-reliability processing across sensor-BS-actuator links, necessitating a joint design of communication, computation, and control. This is challenging due to the coupling of communication and computation in the closed-loop system, where sensing data latency and reliability directly impact command signal computation. Moreover, subsystems sharing distributed BS resources face load imbalances, leading to inefficient resource use and degraded performance. Efficient sensor-actuator association and resource allocation are thus essential to balance loads and ensure reliable subsystem operation within limited resources.

In the literature, there have been several works investigating wireless networked control with joint communication and control co-design while ignoring the computation process \cite{control1, communication1, cod6}.
For instance, \cite{control1} studied a wireless networked control system with a single BS coordinating multiple subsystems.
The works \cite{communication1} focused on the optimization of communication under control stability constraints to minimize energy consumption. 
Furthermore, \cite{ cod6} explored the joint communication and control co-design in wireless networked control systems, optimizing network energy consumption and control costs.
On the other hand, there is another line of related work investigating the joint communication and computation design in MEC networks to enhance the computation performance \cite{MEC1, MEC2, li2023joint}.
In particular, \cite{MEC1, MEC2} focused on optimizing the communication and computation resource allocation in MEC networks to improve the energy efficiency and reduce the computation latency. \cite{MEC1} studied a single-cell MEC system with multiple users, in which the system energy consumption was minimized while meeting the users' computation latency requirements.
\cite{MEC2} further considered the multi-cell MEC systems, in which the user association becomes a new design issue for enhancing the overall system performance.    
Despite the advancements in communication-control co-design and communication-computation co-design independently, there is only a handful of prior works studying the co-design of control, communication, and computation processes in MEC-enabled wireless networked control systems\cite{li2023joint}.
However, these prior works focused on the scenario with a single BS, which may not work well for practical large-scale scenarios, in which multiple BSs are deployed to support a large number of connected devices in IIoT over a large area.

Different from prior works, this paper considers a wireless networked control system with multiple BSs, each equipped with an MEC server and a controller, for coordinating the control of multiple subsystems comprising plants, sensors, and actuators. The distributed communication and computation resources among BSs introduce new challenges, including BS-sensor/actuator association, communication-computation coordination, and handling inter-subsystem interference. To address these, we propose a TDMA protocol for the cascaded communication and computation process, where uplink data collection and downlink command broadcasting occur over orthogonal time slots. Massive MIMO at BSs enables serving multiple sensors and actuators via spatial multiplexing. We jointly optimize BS-sensor/actuator associations and resource allocation to minimize closed-loop control latency while accounting for HRLLC uncertainties and inter-user interference. To solve the non-convex problem, we employ alternating optimization and SCA techniques. Numerical results show the proposed design outperforms benchmark schemes and demonstrates the superiority of TDMA over FDMA, especially under limited computation resources.

\section{System Model}
\begin{figure}[!htbp]
\centering
  \includegraphics[width=3.6 in]{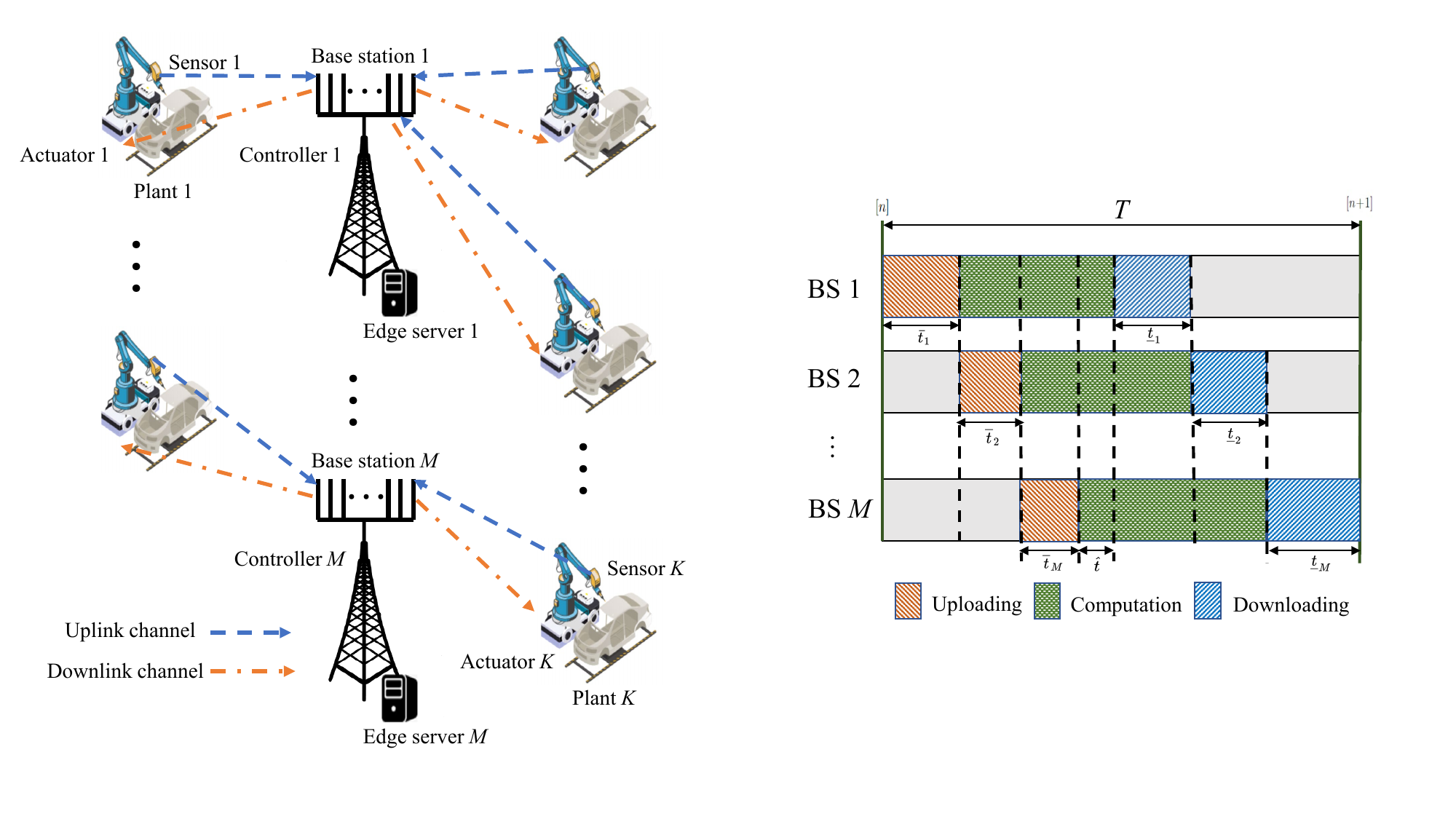}
\caption{The MEC-enabled wireless networked control system with multiple BSs serving a set of control subsystems.TDMA operational protocol for control across multiple BSs.}
\label{fig:TDMA}
\end{figure}
\vspace*{-0.5\baselineskip} 
\vspace*{-0.5\baselineskip} 

As shown in Fig.~\ref{fig:TDMA}, we consider a wireless networked control system, in which $M$ BSs cooperatively coordinate the wireless control of $K$ subsystems each consisting of a plant, a sensor, and an actuator.
Let $\mathcal M = \{1, \ldots, M\}$ denote the set of BSs, and $\mathcal K = \{1, \ldots, K\}$ denote the set of subsystems. 
Each BS is attached to an edge server acting as a controller, and equipped with $N$ antennas.
With massive MIMO employed at the BSs, we assume that $N \gg K$.
Note that in the wireless networked control system, each plant of subsystem $k\in \mathcal K$ is sampled at regular intervals defined by the period $T$, which corresponds to the total duration for one-round processing.
In this work, we focus on one sample period of duration $T$, in which each subsystem (and its plant, sensor, and actuator) is associated with one single BS, and each BS may be associated with multiple subsystems. 
In each sample period, each BS $m\in \mathcal M$ first receives the sensing information from sensors of its associated subsystems and then computes the feedback command.
Subsequently, BS $m$ sends the command signals back to the corresponding actuators of the subsystems for wireless control.
With massive MIMO, each BS can serve its associated sensors or actuators over the same time-frequency resource via spatial multiplexing. 
Let $\{\alpha_{m,k}\}$ denote the BS-sensor/actuator association indicators, which are design variables to be optimized later.
In particular, $\alpha_{m,k}=1$ means that subsystem $k\in \mathcal{K}$ is associated with BS $m\in \mathcal{M}$, and $\alpha_{m,k}=0$ otherwise. 
It should be noted that each subsystem is permitted to be associated with only one BS.
Thus, we have 
\vspace*{-0.5\baselineskip}
\vspace*{-0.3\baselineskip} 
\begin{align}
&\alpha_{m,k} \in\{0,1\}, \quad \forall m \in \mathcal{M}, k \in \mathcal{K}, \\
&\sum_{m \in \mathcal{M}} \alpha_{m,k} = 1, \quad \forall k \in \mathcal{K}.
\end{align}
\vspace*{-0.5\baselineskip}
\vspace*{-0.3\baselineskip} 

Furthermore, we consider the TDMA protocol among BSs as illustrated in Fig.~\ref{fig:TDMA}, in which different BSs serve their respectively associated sensors and actuators over orthogonal time slots.
In particular, the whole sample period of duration \( T \) is divided into \( 2M + 1 \) time slots, in which each of the first \( M \) slots is utilized for sensors' data uploading in one BS, the $(M+1)$-th slot is reserved for computation, and each of the last \( M \) slots is used for commands downloading in one BS.
Specifically, we denote the set of sensors/actuators  associated with each BS $m$ as $\mathcal K_m$, in which $\bigcup_{m=1}^M \mathcal K_m = \mathcal K = \{1,...,K\}$. 
At each slot $m\in \{1,..., M\}$, the sensors of plant $k\in \mathcal K_m$ first transmit its state vector $\hat{\boldsymbol{x}}_{k}\in\mathbb{R}^{a\times 1}$  to its associated BS $m$, where $a$ represents the number of state parameters.
The short-packet communication is considered to meet the stringent latency requirements.
From slot $(m+1)$ to slot $(M+m)$, BS $m$ processes the received sensing data $\{\hat{\boldsymbol{x}}_{k}\}_{k\in \mathcal K_m}$ to obtain the corresponding control command signal $\hat{\mathbf{u}}_{k}\in\mathbb{R}^{a\times 1}$ for each sub-system $k\in \mathcal K_m$ via edge computing.
Subsequently, in slot $M+m+1$, BS $m$ sends the control command signal $\hat{\mathbf{u}}_{k}\in\mathbb{R}^{a\times 1}$ back to plant $k\in \mathcal K_m$ for remote control. 
Let $\overline{\lambda}_{k}$ (in bits) denote the size of state vector $\hat{\boldsymbol{x}}_{k}$  for uplink transmission of plant $k$, $\hat{\lambda}_{k}$ denote the size of input bits for edge computing, and $\underline{\lambda}_{k}$ denote the size of control signal $\hat{\mathbf{u}}_{k}$ for downlink transmission.
For BS $m \in \mathcal{M}$, we denote $\overline{t}_{m}$ as the duration of slot $m$ for uploading, $\hat{t}$ as the duration of slot $M+1$ for dedicated computation, and $\underline{t}_{m}$ as the duration of slot $M+m+1$ for commands downloading.
Therefore, the total sample duration $T$ is given by $T=\sum_{m=1}^M \bar{t}_m+\hat{t}+\sum_{m=1}^M \underline{t}_m$, which is also a design variable to be optimized.
\vspace*{-0.5\baselineskip} 
\vspace*{-0.5\baselineskip} 
\subsection{Wireless Control Model}
We first introduce the wireless networked control model\cite{bishop2011modern} by focusing on one particular subsystem $k\in \mathcal K$.
In general, the state vector $\hat{\boldsymbol{x}}_{k}(t)$ of subsystem $k$ changes continuously with respect to time $t$.
Let $\mathbf{A}_{k}\in \mathbb{R}^{a\times a}$ and $\mathbf{B}_{k}\in \mathbb{R}^{a\times 1}$ denote the system parameters for plant $k$, and $\mathbf{n}_{k}(t)\in \mathbb{R}^{a\times 1}$ denote the disturbance or the independent and identically distributed  additive white Gaussian noise (AWGN) with mean zero and covariance $\mathbf{R}_{k}$.
The change of state vector $\hat{\boldsymbol{x}}_{k}(t)$ of subsystem $k$ is governed by 
$d \hat{\mathbf{x}}_{k}(t)=\mathbf{A}_{k} \hat{\mathbf{x}}_{k}(t) d t+\mathbf{B}_{k} \hat{\mathbf{u}}_{k}(t) d t+d \mathbf{n}_{k}(t)$.
Assume that the sampling of each plant’s state vectors occurs at the beginning of each sample period $i$, and the control commands are input at the end of each sample period.
Then, following \cite{li2023joint}, the discrete-time control model is expressed as
$\boldsymbol{x}_{k}[i+1]=\mathbf{\Gamma}_{k} \boldsymbol{x}_{k}[i]+\boldsymbol{\Lambda}_{k} \mathbf{u}_{k}[i]+\mathbf{n}_{k}[i],$
where $\boldsymbol{x}_{k}[i]$ denotes the state vector of sample $i$ while $\mathbf{u}_{k}[i]$ denotes the input vector of sample $i$.
We have $\mathbf{\Gamma}_{k}=e^{T \mathbf{A}_{k}} \in \mathbb{R}^{a \times a}$,\footnote{Here, $e^{T \mathbf{A}_k}=\mathbf{I}+\mathbf{A}_k T+\frac{1}{2}\left(\mathbf{A}_k T\right)^2+\cdots+\frac{1}{n !}\left(\mathbf{A}_k T\right)^n+\cdots$.} and $\mathbf{\Lambda}_{k}=\left(\int_0^T e^{\mathbf{A}_{k} t} d t\right) \mathbf{B}_{k} \in \mathbb{R}^{a \times a}$\cite{park2017wireless}.
We use the Lyapunov-like function to analyze the stability of the system, denoted as $\Delta\left(\boldsymbol{x}_{k}[i]\right)=\boldsymbol{x}_{k}[i]^{T} \mathbf{Q}_{k} \boldsymbol{x}_{k}[i]$, where $\mathbf{Q}_{k}$ is a given positive definite matrix\cite{gatsis2015opportunistic}. 
To ensure the control stability, the Lyapunov-like function $\Delta\left(\boldsymbol{x}_{k}[i]\right)$ should decrease at a given converge rate $\eta_{k}<1$ \cite{gatsis2015opportunistic}.
Due to inherent variability from sensor noise, system perturbations, and communication errors, $\boldsymbol{x}_{k}[i]$ is treated as a stochastic variable.
Therefore, we apply the expectation of $\Delta(\boldsymbol{x}_{k}[i])$ to express the stability condition for the control system as 
\vspace*{-0.5\baselineskip} 
\begin{equation}\label{eq7}
\mathbb{E}\left[\Delta\left(\boldsymbol{x}_{k}[i+1]\right) \mid \boldsymbol{x}_{k}[i]\right] \leq \eta_{k} \Delta\left(\boldsymbol{x}_{k}[i]\right)+\operatorname{Tr}\left(\mathbf{Q}_{k} \mathbf{R}_{k}\right),
\end{equation}
where $\operatorname{Tr}\left(\mathbf{Q}_{k} \mathbf{R}_{k}\right)$ is caused by the randomness of the control system.

Next, we consider the transmission of sensing and  control command signals. 
Packet loss may occur at both the uplink and downlink transmission due to the short-packet transmission, wireless channel fading, and inter-plant interference.
In this case, the control input $\mathbf{u}_{k}[i]$  depends on whether the transmission is successful or not.
Specifically, when both the state data and command signals are transmitted successfully, a linear negative feedback mechanism is employed for closed-loop control, based on which we have $\mathbf{u}_k[i]=-{\mathbf{F}}_k \mathbf{x}_k[i]$, where ${\mathbf{F}}_k$ is the feedback coefficient.
Otherwise, if either transmission fails, we set $\mathbf{u}_k[i]=0$ for open-loop control. 
Let $\boldsymbol{\Gamma}_{k}^{\text {close }}=\boldsymbol{\Gamma}_{k}-\mathbf{\Lambda}_{k} \mathbf{F}_{k}$ denote the closed-loop control parameter, $\boldsymbol{\Gamma}_{k}^{\text {open }}=\boldsymbol{\Gamma}_{k}$ denote the open-loop control parameter, and $\varepsilon_{k}$ denote the overall outage probability of control subsystem $k$.
Here, $\varepsilon_k$ will be specified later based on the short-packet transmission for HRLLC.
Based on the above analysis, the stability condition in (\ref{eq7}) is re-expressed as
\vspace*{-0.5\baselineskip} 
\begin{small}
\begin{equation}\label{control_constraint1}
    \boldsymbol{x}_{k}[i]^{T}\!\!\left[\!\eta_{k}\mathbf{Q}_{k}\!\!-\!\!(1\!\!-\!\!\varepsilon_{k}){\mathbf{\Gamma}_{k}^{\rm{close}}}^{T}\!\mathbf{Q}_{k}\mathbf{\Gamma}_{k}^{\rm{close}}\!\!-\!\!\varepsilon_{k}{\mathbf{\Gamma}_{k}^{\rm{open}}}^{T}\!\mathbf{Q}_{k}\mathbf{\Gamma}_{k}^{\rm{open}}\!\right]\!\boldsymbol{x}_{k}[i]\! \succeq\! \mathbf{0}.
\end{equation}    
\end{small}
Note that the inequality in (\ref{control_constraint1}) must hold for any plant state $\mathbf{x}_k[i]\neq 0$.
Therefore, (\ref{control_constraint1}) is equivalently rewritten as the following semi-definite constraint \cite{li2023joint}: 
\vspace*{-0.5\baselineskip} 
\begin{equation}\label{control_constraint}
    \eta_{k}\mathbf{Q}_{k}-(1-\varepsilon_{k}){\mathbf{\Gamma}_{k}^{\rm{close}}}^{T}\mathbf{Q}_{k}\mathbf{\Gamma}_{k}^{\rm{close}}-\varepsilon_{k}{\mathbf{\Gamma}_{k}^{\rm{open}}}^{T}\mathbf{Q}_{k}\mathbf{\Gamma}_{k}^{\rm{open}} \succeq \mathbf{0}.
\end{equation}

\subsection{Communication between BSs and Plants}
In this subsection, we consider the wireless transmission for state information uploading from the plants to the corresponding BS in the uplink and that for control command downloading in the downlink.
Let $\boldsymbol{g}_{m,k}\in \mathbb{C}^{N \times 1}$ denote the channel vector between plant $k$ and BS $m$. 
Note that the massive MIMO is employed at each BS to serve its associated plants via spatial multiplexing.
In particular, the BS implements the low-complexity matched-filter (MF)  based linear receiver by setting the receive beamforming vector as  $\bar{\boldsymbol{w}}_{m,k}=\frac{{\boldsymbol{g}_{m,k}^\textit{H}}}{\left\|\boldsymbol{g}_{m,k}\right\|}$ for its associated plant $k$.\footnote{Note that the MF based receiver is asymptotically optimal when $M$ becomes sufficiently large and the multi-antenna channels are independently distributed, thus making it a practically appealing choice \cite{remove}.}
Let $\bar p_{m,k}$ denote the transmission power of plant $k$ during the uploading.
Let $\xi_{m,l,k}=\left|{\bar{\boldsymbol{g}}_{m,l}^\textit{H}\bar{\boldsymbol{w}}_{m,k}} \right|^2$ denote the equivalent channel power gain from plant $l$ to BS $m$ based on the MF-based receive beamforming for plant $k$.
Accordingly, the received signal-to-interference-plus-noise ratio (SINR) at the BS for its associated plant $k$ is expressed as 
\vspace*{-0.5\baselineskip} 
\begin{equation}
\label{sinr_ul}
\bar \gamma_{m,k}=\frac{\bar p_{m,k}\xi_{m,k,k}}{\sum_{l=1,l\neq k}^{K} \bar p_{m,l}\xi_{m,l,k}+\sigma^2},
\end{equation}
where $\sigma^2$ denotes the noise power at the BS receiver.
Due to the HRLLC requirements, we consider the short-packet or finite blocklength transmission.
We denote $B_0$ as the system bandwidth, and $\bar{\varepsilon}_{k}$ as the outage probability for the sensor of plant $k$ uploading data to BS $m$.
In such a case, according to \cite{polyanskiy2010channel}, the achievable uplink transmission rate for plant $k$ is expressed as 
\vspace*{-0.5\baselineskip}
\begin{small}
\begin{equation}
\label{fbl}
\bar{r}_{k}\!\approx\! B_0\!\left[\!\sum_{m=1}^{M}\alpha_{m,k}\log_2 \left(\!1\!+\!\bar{\gamma}_{m,k}\!\right)\! - \!\sqrt{\frac{\tilde{V}}{\sum_{m=1}^{M}\alpha_{m,k}\bar{t}_{m} B_0}} Q^{-1}\left(\bar{\varepsilon}_{k}\right)\!\right]\!,
\end{equation}
\end{small}
where $Q^{-1}(x)$ denotes the inverse of Q-function defined as $Q(x)=\int_x^{\infty} \frac{1}{\sqrt{2 \pi}} e^{-\frac{t^2}{2}} d t$, and $\tilde{V}=(\log_2 e)^2$ denotes the approximated channel dispersion\cite{yang2014quasi}.
Based on (\ref{fbl}), the uplink outage probability $\bar{\varepsilon}_{k}$ is expressed as 
\vspace*{-0.5\baselineskip} 
\begin{small}
\begin{equation}
\label{err_uplink}
\bar{\varepsilon}_{k}=Q\!\left(\!\log_e 2 \cdot \sum_{m=1}^{M}\!\alpha_{m,k}\!\left(\!\sqrt{\bar{t}_{m} B_0}\log_2\!\left(1+\bar{\gamma}_{m,k}\right)\!-\!\frac{\bar{\lambda}_{k}}{\sqrt{\bar{t}_{m} B_0}}\right)\!\right).
\end{equation}
\end{small}
\vspace*{-0.5\baselineskip} 

Next, we consider the command signals broadcasting in the last $M$ slots after the BSs obtain the control command signals.
Each BS employs the massive MIMO with MF precoding to transmit these control signals.
Let $\zeta_{m,l,k}=\left|{\underline{\boldsymbol{g}}_{m,k}}^\textit{H}\underline{\boldsymbol{w}}_{m,l}\right|^2$ denote the equivalent power gain from BS $m$ to the actuator of plant $k$, where $\underline{\boldsymbol{w}}_{m,l}=\frac{{\boldsymbol{g}_{m,l}^\textit{H}}}{\left\|\boldsymbol{g}_{m,l}\right\|}$ is the precoding vector for plant $l$.
Let $\underline p_{m,k}$ denote the downlink transmission power of BS $m$ for plant $k$.
Let $\underline{P}_m$ denote the total downlink transmission power available at BS $m$ for all associated plants.
Therefore, we have $\sum_{k \in \mathcal{K}} \alpha_{m, k} \underline p_{m,k} \leq \underline{P}_m, \forall m \in \mathcal{M}.$ 
Accordingly,  the SINR at the receiver of plant $k$ from its associated BS $m$ during the downlink process is expressed as
\vspace*{-0.5\baselineskip} 
\begin{equation}
\underline \gamma_{m,k}=\frac{\underline p_{m,k}\zeta_{m,k,k}}{\sum_{l=1,l\neq k}^{K}\underline p_{m,l}\zeta_{m,l,k}+\sigma^2},
\label{sinr_down}
\end{equation}
Similar to (\ref{err_uplink}), the downlink outage probability is expressed as
\vspace*{-0.5\baselineskip} 
\begin{small}
\begin{equation}
\label{err_downlink}
\underline{\varepsilon}_{k}=Q\!\left(\!\log_e 2 \cdot {\sum_{m=1}^{M}\alpha_{m,k}}\!\left(\!\sqrt{\underline{t}_{m} B_0} \log_2 \left(1+\underline{\gamma}_{m,k}\!\right)\!-\!\frac{\underline{\lambda}_{k}}{\sqrt{\underline{t}_{m} B_0}}\!\right)\!\right).
\end{equation}    
\end{small}

Furthermore, the transmission is successful only if there is no outage occured in both uplink and downlink.
Therefore, the overall outage probability $\varepsilon_{k}$ of plant $k$ is expressed as $\varepsilon_{k}=1-(1-\bar{\varepsilon}_{k})(1-\underline{\varepsilon}_{k}).$

\subsection{Edge Computation at BSs}
As shown in Fig.~\ref{fig:TDMA}, after the state information is uploaded from the subsystems, the corresponding BS starts the edge computation immediately to obtain the feedback control information for the subsystems.
We use $F_{m}$ to denote the maximum computation frequency at BS $m$, and $S_{k}$ (in cycles/bit) to denote the CPU cycles required for computing one bit of the task from subsystem $k$. Recall that $\hat{\lambda}_{k}$ is used to denote the number of computing bits for plant $k$.
Following the TDMA protocol, we have the following constraints to ensure that the computing tasks are successfully completed within the allocated time periods (i.e., from $(m+1)$ to slot $(M+m)$):
\vspace*{-0.5\baselineskip} 
\begin{equation}
 \sum_{k \in \mathcal{K}}\alpha_{m,k} S_{k} \hat{\lambda}_{k} \leq F_{m}\left(\sum_{i=m+1}^M \bar{t}_i + \hat{t} + \sum_{i=1}^{m-1} \underline{t}_i\right), \forall m \in \mathcal{M},
 \label{16}
\end{equation}
where the right-hand-side represents the total computational capacity available at BS $m$ within the allocated time slots.
This takes into account the durations of the uploading time slots of the subsequent BSs $\left(\sum_{i=m+1}^M \bar{t}_i\right)$, the dedicated computation time $\hat{t}$, and the downloading time slots of the preceding BSs $\left(\sum_{i=1}^{m-1} \underline{t}_i\right)$. 
The left-hand-side of (\ref{16}) represents the total computational load required by BS $m$ to process the computing bits for all associated subsystems.

\subsection{Problem Formulation}
Our objective is to minimize the sample period $T$ of the whole network under the constraints on control stability, maximum transmission power, and communication error probability.
The rationale behind minimizing $T$ lies in the fact that a shorter sampling period enhances the system's ability to converge more rapidly to a stable state, thereby improving the overall control performance.
This requires a joint management of the BS-sensor/actuator association, transmission power control, and time allocations.

To facilitate the derivation, we first reformulate the control stability constraint (\ref{control_constraint}).
Towards this end, we introduce $\varepsilon_{\text{th}}$ as a threshold for the communication reliability or the maximum outage probability for uploading and downloading. 
Based on this, we consider the most stringent scenario for control stability, by setting $\varepsilon_{k}=1-(1-\varepsilon_{\text{th}})^2$. Accordingly, constraint (\ref{control_constraint}) becomes
\vspace*{-0.5\baselineskip} 
\begin{equation}
\begin{aligned}
&\eta_{k}\mathbf{Q}_{k} - (1-\varepsilon_{\text{th}})^2{\mathbf{\Gamma}_{k}^{\rm{close}}}^{T}\mathbf{Q}_{k}\mathbf{\Gamma}_{k}^{\rm{close}} \\
&- (1-(1-\varepsilon_{\text{th}})^2){\mathbf{\Gamma}_{k}^{\rm{open}}}^{T}\mathbf{Q}_{k}\mathbf{\Gamma}_{k}^{\rm{open}} \succeq \mathbf{0}, \quad \forall k\in\mathcal{K}.
\label{15}
\end{aligned}
\end{equation}
Additionally, we approximate $\mathbf{\Gamma}_{k}=e^{T \mathbf{A}_{k}}$ and $\mathbf{\Lambda}_{k}=\left(\int_0^T e^{\mathbf{A}_{k} t} d t\right) \mathbf{B}_{k}$ as $\mathbf{I}+T \mathbf{A}_{k}$ and  $T\mathbf{B}_{k}$, respectively.
These simplifications are justified because the sampling time $T$ is sufficiently small such that the higher-order terms are safely ignored \cite{lee1967foundations}.
Based on the above approximations, constraint (\ref{control_constraint}) or (\ref{15}) is further approximated as 
\vspace*{-0.5\baselineskip} 
\begin{IEEEeqnarray}{l}
\left(\eta_{k}-1\right) \mathbf{Q}_{k} + \left[\left(1-\varepsilon_{\mathrm{th}}\right)^2\left(\mathbf{K}_{k}^{T} \mathbf{B}_{k}^{T} \mathbf{Q}_{k} + \mathbf{Q}_{k} \mathbf{B}_{k} \mathbf{K}_{k}\right) \right. \nonumber\\
\left. - \left(\mathbf{A}_{k}^{T} \mathbf{Q}_{k} + \mathbf{Q}_{k} \mathbf{A}_{k}\right)\right]T \nonumber \\
+ \left[\left(1-\varepsilon_{\mathrm{th}}\right)^2\left(\mathbf{A}_{k}^{T} \mathbf{Q}_{k} \mathbf{B}_{k} \mathbf{K}_{k} + \mathbf{K}_{k}^{T} \mathbf{B}_{k}^{T} \mathbf{Q}_{k} \mathbf{A}_{k} \right. \nonumber\right. \\
\left.\left. - \mathbf{K}_{k}^{T} \mathbf{B}_j^{T} \mathbf{Q}_j \mathbf{B}_j \mathbf{K}_j\right) - \mathbf{A}_{k}^{T} \mathbf{Q}_{k} \mathbf{A}_{k}\right] T^2 \succeq 0, \quad \forall k \in \mathcal{K}.
\label{eqn:example}
\IEEEeqnarraynumspace
\end{IEEEeqnarray}
Furthermore, we define
\begin{small}
\begin{equation}
\mathbf{\Upsilon}_k =\left[\left(1\!-\!\varepsilon_{\mathrm{th}}\right)^2\!\left(\!\mathbf{K}_{k}^{T} \mathbf{B}_{k}^{T} \mathbf{Q}_{k}\!+\! \mathbf{Q}_{k} \mathbf{B}_{k} \mathbf{K}_{k}\!\right)\!- \!\left(\!\mathbf{A}_{k}^{T} \mathbf{Q}_{k} \!+\! \mathbf{Q}_{k} \mathbf{A}_{k}\!\right)\!\right],
\end{equation}    
\end{small}
\begin{equation}
\begin{aligned}
\mathbf{\Phi}_k &= \biggl[\left(1-\varepsilon_{\mathrm{th}}\right)^2\biggl(\mathbf{A}_{k}^{T} \mathbf{Q}_{k} \mathbf{B}_{k} \mathbf{K}_{k} + \mathbf{K}_{k}^{T} \mathbf{B}_{k}^{T} \mathbf{Q}_{k} \mathbf{A}_{k}\biggr) \\
& - \left(1-\varepsilon_{\mathrm{th}}\right)^2 \mathbf{K}_{k}^{T} \mathbf{B}_j^{T} \mathbf{Q}_j \mathbf{B}_j \mathbf{K}_j - \mathbf{A}_{k}^{T} \mathbf{Q}_{k} \mathbf{A}_{k}\biggr],
\end{aligned}
\end{equation}
and re-express (\ref{control_constraint}) or (\ref{eqn:example}) as 
\vspace*{-0.5\baselineskip} 
\begin{equation}
\mathbf{\Phi}_k T^2+\mathbf{\Upsilon}_kT+\left(\eta_{k}-1\right) \mathbf{Q}_{k} \succeq \mathbf{0}, \forall k \in \mathcal{K},
\label{control_re}
\end{equation}
where $\mathbf{\Phi}_k$, $\mathbf{\Upsilon}_k$, and $\mathbf{Q}_{k}$ are real symmetric matrices. 

For notational convenience, we denote $\boldsymbol{\alpha} \in \mathbb{R}^{M \times K}$ as a matrix representing the BS-sensor/actuator association, with $[\boldsymbol{\alpha}]_{m,k} = \alpha_{m,k}$, denote $\boldsymbol{p} = \{\bar{p}_{m,k}, \underline{p}_{m,k}\}$ as the transmission power variable, and denote $\boldsymbol{t} = \{\bar{t}_m, \hat{t}, \underline{t}_m, T\}$ as the time allocation variables to be optimized.
As such, the control stability constrained latency minimization problem is formulated as 
\vspace*{-0.5\baselineskip} 
\begin{align}
\text{(P1)}&:\underset{\substack{\boldsymbol{p},\boldsymbol{t},\boldsymbol{\alpha}}}{\text{min}}\quad T \label{Problem}\\
\text{s.t.}
&0\leq \bar{\varepsilon}_{k} (\alpha_{m,k}, \bar p_{m,k}, \bar{t}_m) \leq {\varepsilon}_{\text{th}},  \forall m\in\mathcal M,  k\in\mathcal K\tag{\ref{Problem}{a}} \label{Probleme},\\
&0\leq \underline{\varepsilon}_{k} (\alpha_{m,k}, \underline p_{m,k}, \underline{t}_m) \leq {\varepsilon}_{\text{th}},  \forall m\in\mathcal M,  k\in\mathcal K\tag{\ref{Problem}{b}} \label{Problemf},\\
&\mathbf{\Phi}_k T^2+\mathbf{\Upsilon}_kT+\left(\eta_{k}-1\right) \mathbf{Q}_{k} \succeq \mathbf{0}, \forall k\in\mathcal K\tag{\ref{Problem}{c}} \label{Problemg},\\
&F_{m}\!\left(\sum_{i=m+1}^M\! \bar{t}_i\!+\!\hat{t}\!+\!\sum_{i=1}^{m-1} \underline{t}_i\right)\!\ge \!\sum_{k \in \mathcal{K}}\alpha_{m,k}S_{k}\hat{\lambda}_{k}, \forall m\in \mathcal{M}\tag{\ref{Problem}{d}} \label{Problemh},\\
&\sum_{m=1}^M \bar{t}_m+\hat{t}+\sum_{m=1}^M \underline{t}_m=T\tag{\ref{Problem}{e}}, \label{Problemi}\\
&\alpha_{m,k} \in\{0,1\},  \forall m \in \mathcal{M}, k \in \mathcal{K}\tag{\ref{Problem}{f}} \label{Problema},\\
&\sum_{m \in \mathcal{M}} \alpha_{m,k} = 1,  \forall k \in \mathcal{K}  \tag{\ref{Problem}{g}} \label{Problemb},\\
&\alpha_{m,k} \bar p_{m,k}\leq \bar{P}_k,  \forall m \in \mathcal{M}, k \in \mathcal{K} \tag{\ref{Problem}{h}} \label{Problemc},\\
&\sum_{k \in \mathcal{K}} \alpha_{m, k} \underline p_{m,k} \leq \underline{P}_m,  \forall m \in \mathcal{M} \tag{\ref{Problem}{i}} \label{Problemd}.
\end{align}
In problem (P1), constraints (\ref{Probleme}) and (\ref{Problemf}) ensure the HRLLC transmission for uplink and downlink.
Constraint (\ref{Problemg}) is the control constraint that ensures the stability of the control system, and constraint (\ref{Problemh}) specifies the computation resource limitations.
Problem (P1) is a non-convex combinatorial optimization problem, which is challenging to solve.

\section{Proposed Solution to Problem (P1)}
In this section, we first relax the binary variable constraints in (\ref{Problema}) into continuous forms as
\vspace*{-0.5\baselineskip} 
 \begin{equation}
     \alpha_{m,k} \in[0,1], \quad \forall m \in \mathcal{M}, k \in \mathcal{K}.
     \label{alpha_relax}
 \end{equation} 
Accordingly, problem (P1) is relaxed as
\vspace*{-0.5\baselineskip} 
\begin{align}
\text{(P1.1)}:~\underset{\boldsymbol{p},\substack{\boldsymbol{t}, \boldsymbol{\alpha}}}{\text{min}}&\quad T \label{Problem_1.1}\\
\text {s.t.}
&~ \text{(\ref{Probleme}), (\ref{Problemf}), (\ref{Problemg}), (\ref{Problemh}), (\ref{Problemi}), (\ref{Problemb}), }\notag\\
&~\text{(\ref{Problemc}), (\ref{Problemd}), \text{and} (\ref{alpha_relax}).}\notag
\end{align}
We will handle problem (P1.1) accordingly next by using the alternating optimization, in which $\{\boldsymbol{t}, \boldsymbol{p}\}$ and $\{\boldsymbol{t}, \boldsymbol{\alpha}\}$ are optimized under given $\boldsymbol{\alpha}$ and $\boldsymbol{p}$, respectively, in an alternating manner.
First, we optimize $\boldsymbol{\alpha}$ and $\boldsymbol{t}$ in problem (P1.1) with given $\boldsymbol{p}$.

\subsection{Optimization of $\boldsymbol{t}$ and $\boldsymbol{\alpha}$ with Given $\boldsymbol{p}$}
The optimization problem of $\boldsymbol{t}$ and $\boldsymbol{\alpha}$ is formulated as
\vspace*{-0.5\baselineskip} 
\begin{align}
\text{(P2)}:~\underset{\substack{\boldsymbol{t}, \boldsymbol{\alpha}}}{\text{min}}&\quad T \label{Problem_ta}\\
\text {s.t.}
&~0\!\leq\! \bar{\varepsilon}_{k} (\alpha_{m,k}, \bar{t}_m) \!\leq\! {\varepsilon}_{\text{th}}, \forall m\in\mathcal M,  k\in\mathcal K\tag{\ref{Problem_ta}{a}} \label{Problema_ta},\\
&~0\!\leq\! \underline{\varepsilon}_{k} (\alpha_{m,k}, \underline{t}_m)\! \leq\! {\varepsilon}_{\text{th}}, \forall m\in\mathcal M, k\in\mathcal K\tag{\ref{Problem_ta}{b}} \label{Problemb_ta},\\
&~ \text{(\ref{Problemg}), (\ref{Problemh}), (\ref{Problemi}), (\ref{Problemb}), (\ref{Problemc}), \text{and} (\ref{alpha_relax}).}\notag
\end{align}
Problem \text{(P2)} is non-convex since the reliability constraints in (\ref{Problema_ta}) and (\ref{Problemb_ta}) and control stability constraint (\ref{Problemg}) are non-convex.
Thus, we introduce auxiliary variables to deal with them.

First, based on the outage probability formulas in (\ref{err_uplink}) and (\ref{err_downlink}), the transmission reliability constraints (\ref{Problema_ta}) and (\ref{Problemb_ta}) are respectively re-expressed as 
\vspace*{-0.5\baselineskip} 
\begin{small}
\begin{equation}
\begin{aligned}
\log_e 2 \cdot \sum_{m=1}^{M}\!\alpha_{m,k}\!\left(\!\sqrt{\bar{t}_{m} B_0} \log_2 \left(1+\bar{\gamma}_{m,k}\right)\!-\!\frac{\bar{\lambda}_{k}}{\sqrt{\bar{t}_{m} B_0}}\!\right)
\!\geq\! Q^{-1}\left({\varepsilon}_{\text{th}}\right),
\label{Problema_ta_re}
\end{aligned}
\end{equation}
\end{small}
\vspace*{-0.5\baselineskip} 
\begin{small}
\begin{equation}
\begin{aligned}
\log_e 2 \cdot \sum_{m=1}^{M}\!\alpha_{m,k}\!\left(\!\sqrt{\underline{t}_{m} B_0} \log_2 \left(1+\underline{\gamma}_{m,k}\right)\!-\!\frac{\underline{\lambda}_{k}}{\sqrt{\underline{t}_{m} B_0}}\!\right)
\!\geq\! Q^{-1}\!\left({\varepsilon}_{\text{th}}\right).
\label{Problemb_ta_re}
\end{aligned}
\end{equation}

\end{small}
Then, to deal with the non-convex constraints in (\ref{Problema_ta_re}) and (\ref{Problemb_ta_re}),  we introduce auxiliary variables $\boldsymbol{\overline B}\in \mathbb{R}^{M \times K}$  and $\boldsymbol{\underline B}\in \mathbb{R}^{M \times K}$, in which $[\boldsymbol{\overline B}]_{m,k}=\bar\beta_{m,k}$ and $[\boldsymbol{\underline B}]_{m,k}=\underline\beta_{m,k}$, $m\in \mathcal M, k\in \mathcal K$.
Thus, we re-express the communication reliability constraints in (\ref{Problema_ta}) and (\ref{Problemb_ta}) (or (\ref{Problema_ta_re}) and (\ref{Problemb_ta_re})) as the following constraints:
\vspace*{-0.5\baselineskip} 
\begin{equation}
\left(\sqrt{\bar{t}_{m} B_0} \log_2 \left(1+\bar{\gamma}_{m,k}\right)-\frac{\bar{\lambda}_{k}}{\sqrt{\bar{t}_{m} B_0}}\right)\geq \bar\beta_{m,k}, 
\label{aux1}
\end{equation}
\begin{equation}
\left(\sqrt{\underline{t}_{m} B_0} \log_2 \left(1+\underline{\gamma}_{m,k}\right)-\frac{\underline{\lambda}_{k}}{\sqrt{\underline{t}_{m} B_0}}\right)\geq \underline\beta_{m,k},
\label{aux2}
\end{equation}
\begin{equation}
    \log_e 2 \cdot {\sum_{m=1}^{M}\alpha_{m,k}}\bar\beta_{m,k} \geq Q^{-1}\left({\varepsilon}_{\text{th}}\right),
\label{27}
\end{equation}
\begin{equation}
    \log_e 2 \cdot {\sum_{m=1}^{M}\alpha_{m,k}}\underline\beta_{m,k} \geq Q^{-1}\left({\varepsilon}_{\text{th}}\right).
\label{28}
\end{equation}
Furthermore, to decouple the product of optimization variables $\alpha_{m,k}$ and $\bar\beta_{m,k}$, we have $\alpha_{m,k}\bar\beta_{m,k}=\frac{\left(\alpha_{m,k}+\bar\beta_{m,k}\right)^2-\left(\alpha_{m,k}-\bar\beta_{m,k}\right)^2}{4}$.
Similarly, to decouple the product of $\alpha_{m,k}$ and $\underline\beta_{m,k}$, we have
$\alpha_{m,k}\underline\beta_{m,k}=\frac{\left(\alpha_{m,k}+\underline\beta_{m,k}\right)^2-\left(\alpha_{m,k}-\underline\beta_{m,k}\right)^2}{4}$.
As such, we further rewrite (\ref{27}) and (\ref{28}) as 
\vspace*{-0.5\baselineskip} 
\begin{equation}
\begin{aligned}
\label{a1}
\frac{\sum\limits_{m=1}^{M}\!\left(\alpha_{m,k}\!+\!\bar\beta_{m,k}\right)^2\!\!-\!\left(\alpha_{m,k}\!-\!\bar\beta_{m,k}\right)^2}{4\log_2 e}\!\geq\! Q^{-1}\left({\varepsilon}_{\text{th}}\right),\forall k\in\mathcal K,
\end{aligned}
\end{equation}
\begin{equation}
\begin{aligned}
\label{a2}
\frac{\sum\limits_{m=1}^{M}\!\left(\alpha_{m,k}\!+\!\underline\beta_{m,k}\right)^2\!\!-\!\left(\alpha_{m,k}\!-\!\underline\beta_{m,k}\right)^2}{4\log_2 e}\!\geq \!Q^{-1}\left({\varepsilon}_{\text{th}}\right),\forall k\in\mathcal K.
\end{aligned}
\end{equation}   

Next, we deal with the quadratic term in control stability constraint (\ref{Problemg}).
By introducing an auxiliary variable $c$ with $T^2 = c$, we re-express constraint (\ref{Problemg}) as the following  constraints: 
\vspace*{-0.5\baselineskip} 
\begin{equation}
\label{au1}
    T^2 \geq c,
\end{equation}
\vspace*{-0.5\baselineskip} 
\begin{equation}
\label{au2}
    T^2 \leq c,
\end{equation}
\vspace*{-0.5\baselineskip} 
\begin{equation}
\label{a_iteration}
    \mathbf{\Phi}_k c+\mathbf{\Upsilon}_kT+\left(\eta_{k}-1\right) \mathbf{Q}_{k} \succeq \mathbf{0}, \forall k \in \mathcal{K}.
\end{equation}

Finally, by combining the above transformations, problem (P2) is equivalently re-expressed as
\vspace*{-0.5\baselineskip} 
\begin{align}
\text{(P3)}&:~\underset{\substack{\boldsymbol{t}, \boldsymbol{\alpha}, \boldsymbol{\overline{B}}, \boldsymbol{\underline{B}}, c}}{\text{min}}\quad T \label{Problem_pt_appro}\nonumber\\
\text{s.t.} &~ \text{(\ref{Problemb}), (\ref{Problemc}), (\ref{Problemh}), (\ref{Problemi}), (\ref{alpha_relax}), (\ref{aux1}), (\ref{aux2}),}\nonumber\\
&~ \text{(\ref{a1}), (\ref{a2}), (\ref{au1}), (\ref{au2}), \text{and} (\ref{a_iteration}).}\nonumber
\end{align}

For problem (P3), the constraints in (\ref{a1}), (\ref{a2}), and (\ref{au1}) are still non-convex.
Therefore, we further employ the SCA technique to deal with the above constraints in an iterative manner.
Specifically, in each inner iteration $\bar{o}\geq1$, let $T^{(\bar{o})}$, $\alpha_{m,k}^{(\bar{o})}$, $\bar\beta_{m,k}^{(\bar{o})}$, and $\underline\beta_{m,k}^{(\bar{o})}$denote the local points of $T$, $\alpha_{m,k}$, $\bar\beta_{m,k}$, and $\underline\beta_{m,k}$, respectively.
Based on the first-order Taylor expansion, the non-convex constraints in (\ref{a1}) and (\ref{a2}) are expressed as (\ref{aux_11}) and (\ref{aux_22}) at the top of this page, respectively, and (\ref{au1}) is expressed as 
\vspace*{-0.5\baselineskip} 
\begin{figure*}[pt!]
\normalsize
\setcounter{MYtempeqncnt}{\value{equation}}
\setcounter{equation}{36}
\begin{equation}
\begin{aligned}
{\sum_{m=1}^{M}\!\left(\!\!\left(\!\alpha_{m,k}^{(\bar{o})}\!+\!\bar\beta_{m,k}^{(\bar{o})}\!\right)^2\!\!+\!2\!\left(\!\alpha_{m,k}\!-\!\alpha_{m,k}^{(\bar{o})}\!\right)\!\!\left(\!\alpha_{m,k}^{(\bar{o})}\!+\!\bar\beta_{m,k}\!\right)\!+\!2\!\left(\!\bar\beta_{m,k}\!-\!\bar\beta_{m,k}^{(\bar{o})}\!\right)\!\!\left(\!\alpha_{m,k}\!+\!\bar\beta_{m,k}^{(\bar{o})}\!\right)\!\!\right)}\!-\!{\left(\!\alpha_{m,k}\!-\!\bar\beta_{m,k}\!\right)^2}
\!\geq \! {4\log_2 e} \cdot Q^{-1}\!\left(\!{\varepsilon}_{\text{th}}\!\right)\!,\!\forall k\!\in\!\mathcal K,    
\end{aligned}
\label{aux_11}
\end{equation}
\begin{equation}
\begin{aligned}
{\sum_{m=1}^{M}\!\left(\!\!\left(\!\alpha_{m,k}^{(\bar{o})}\!+\!\underline\beta_{m,k}^{(\bar{o})}\!\right)^2\!\!+\!2\!\left(\!\alpha_{m,k}\!-\!\alpha_{m,k}^{(\bar{o})}\!\right)\!\!\left(\!\alpha_{m,k}^{(\bar{o})}\!+\!\underline\beta_{m,k}\!\right)\!+\!2\!\left(\!\underline\beta_{m,k}\!-\!\underline\beta_{m,k}^{(\bar{o})}\!\right)\!\!\left(\!\alpha_{m,k}\!+\!\underline\beta_{m,k}^{(\bar{o})}\!\right)\!\!\right)}\!-\!{\left(\!\alpha_{m,k}\!-\!\underline\beta_{m,k}\!\right)^2}
\!\geq\! {4\log_2 e} \cdot Q^{-1}\!\left(\!{\varepsilon}_{\text{th}}\!\right)\!,\!\forall k\!\in\!\mathcal K.  
\end{aligned}
\label{aux_22}
\end{equation}
\hrulefill
\end{figure*}

\begin{equation}
\label{T_approximate}
    -T^{(\bar{o})^2}+c+\left(-2 T^{(\bar{o})}\right)\left(T-T^{(\bar{o})}\right) \leq 0.
\end{equation}
Accordingly, we approximate problem (P3) as (P4.$\bar{o}$) in the $\bar{o}$-th inner iteration for SCA, which is convex and thus can be optimally solved by convex solvers such as CVX\cite{grant2014cvx}.
\vspace*{-0.5\baselineskip} 
\begin{align}
(\text{P4.}\bar{o})&:~\underset{\substack{\boldsymbol{t}, \boldsymbol{\alpha}, \boldsymbol{\overline{B}}, \boldsymbol{\underline{B}}, c}}{\text{min}}\quad T \nonumber\\
\text{s.t.}
&~ \text{(\ref{Problemh}), (\ref{Problemi}), (\ref{Problemb}), (\ref{Problemc}), (\ref{alpha_relax}), (\ref{aux1}), (\ref{aux2})}, \nonumber\\
&~ \text{(\ref{au2}), (\ref{a_iteration}), (\ref{aux_11}), (\ref{aux_22}), \text{and} (\ref{T_approximate}).} \notag
\end{align}

Notice that the first-order approximations in the left-hand-side of (\ref{aux_11}), (\ref{aux_22}), and (\ref{T_approximate}) serve as upper bounds of those in (\ref{a1}), (\ref{a2}), and (\ref{au1}). Therefore, solving the series of problems in ($\text{P4.}\bar{o}$) is ensured to result in monotonically decreasing objective values.
Therefore, by solving problem ($\text{P4.}\bar{o}$) in an iterative manner, a converged solution to problem (P2) or equivalently (P3) is obtained.

\subsection{Joint Optimization of $\boldsymbol{t}$ and $\boldsymbol{p}$ with Given $\boldsymbol{\alpha}$}
Next, we optimize $\boldsymbol{p}$ and $\boldsymbol{t}$ in problem (P1.1) with given user association $\boldsymbol{\alpha}$.
The problem is expressed as
\vspace*{-0.5\baselineskip} 
\begin{align}
\text{(P5)}:\underset{\substack{\boldsymbol{p}, \boldsymbol{t}}}{\text{min}}&~\quad T \label{Problem_pt} \\
\text {s.t.}
&~0\leq \bar{\varepsilon}_{k} (\bar p_{m,k}, \bar{t}_m) \leq {\varepsilon}_{\text{th}}, \forall m\in\mathcal M,  k\in\mathcal K\tag{\ref{Problem_pt}{a}} \label{Problema_pt},\\
&~0\leq \underline{\varepsilon}_{k} (\underline p_{m,k}, \underline{t}_m) \leq {\varepsilon}_{\text{th}}, \forall m\in\mathcal M, k\in\mathcal K\tag{\ref{Problem_pt}{b}} \label{Problemb_pt},\\
&~\text{ (\ref{Problemg}), (\ref{Problemh}), (\ref{Problemi}), (\ref{Problemc}), \text{and} (\ref{Problemd}).}\notag
\label{Problemg_pt}
\end{align}
Notice that problem \text{(P5)} is non-convex since transmission reliability constraints (\ref{Problema_pt}) and (\ref{Problemb_pt}) as well as control stability constraint (\ref{Problemg}) are non-convex.
Thus, we introduce auxiliary variables to handle them.

First, based on the SINR expressions in (\ref{sinr_ul}) and (\ref{sinr_down}) as well as  the corresponding outage probabilities in (\ref{err_uplink}) and (\ref{err_downlink}), the communication reliability constraints in (\ref{Problema_pt}) and (\ref{Problemb_pt}) are rewritten as 
\vspace*{-0.5\baselineskip} 
\begin{equation}
\begin{aligned}
  &\log_2 e~ Q^{-1}\left(\varepsilon_{\mathrm{th}}\right)+\sum_{m=1}^{M}\alpha_{m,k}\left(\frac{\bar{\lambda}_k}{\sqrt{\bar{t}_m B_0}}\right)\leq\sum_{m=1}^{M}\\&\alpha_{m,k}{\sqrt{\bar{t}_m B_0}}\log _2\left(1+\frac{\bar p_{m,k}\xi_{m,k}}{\sum_{i=1,i\neq k}^{K} \bar p_{m,i}\xi_{m,i,k}+\sigma^2}\right) ,   
\end{aligned}
\end{equation}
\begin{equation}
\begin{aligned}
  &\log_2 e~ Q^{-1}\left(\varepsilon_{\mathrm{th}}\right)+\sum_{m=1}^{M}\alpha_{m,k}\left(\frac{\underline{\lambda}_k}{\sqrt{\underline{t}_m B_0}}\right)\leq\sum_{m=1}^{M}\\&\alpha_{m,k}{\sqrt{\underline{t}_m B_0}}\log _2\left(1+\frac{\underline p_{m,k}\xi_{m,k}}{\sum_{i=1,i\neq k}^{K} \underline p_{m,i}\xi_{m,i,k}+\sigma^2}\right) .
\end{aligned}
\end{equation}
Next, since variables $\boldsymbol{p}$ and $\boldsymbol{t}$ are coupled in constraints (\ref{Problema_pt}) and (\ref{Problemb_pt}), we introduce auxiliary variables $\boldsymbol{\overline D}\in \mathbb{R}^{M \times K}$  and $\boldsymbol{\underline D}\in \mathbb{R}^{M \times K}$, 
in which $[\boldsymbol{\overline D}]_{m,k}=\bar d_{m,k}$ and $[\boldsymbol{\underline D}]_{m,k}=\underline d_{m,k}$.
Thus, we have
\vspace*{-0.5\baselineskip} 
\begin{equation}
    \label{aux_dup}
    \overline d_{m,k}\leq\log _2\left(1+\frac{\bar p_{m,k}\xi_{m,k}}{\sum_{i=1,i\neq k}^{K} \bar p_{m,i}\xi_{m,i,k}+\sigma^2}\right),   
\end{equation}
\begin{equation}
    \label{aux_ddown}
    \underline d_{m,k}\leq\log _2\left(1+\frac{\underline p_{m,k}\xi_{m,k}}{\sum_{i=1,i\neq k}^{K} \underline p_{m,i}\xi_{m,i,k}+\sigma^2}\right). 
\end{equation}
Furthermore, we handle the non-convex constraints in (\ref{aux_dup}) and (\ref{aux_ddown}) by rewriting them as
\vspace*{-0.5\baselineskip} 
\begin{equation}
\label{upd_aux}
\begin{aligned}
 \overline d_{m,k}&\leq \log_2\left(\sum_{i=1}^{K} \bar p_{m,i}\xi_{m,i,k}+\sigma^2\right)
\\&-\log_2\left(\sum_{i=1,i\neq k}^{K} \bar p_{m,i}\xi_{m,i,k}+\sigma^2\right),\forall m\in\mathcal M, k\in\mathcal K,   
\end{aligned}
\end{equation}
\begin{equation}
\label{downd_aux}
\begin{aligned}
\underline d_{m,k}&\leq \log_2\left(\sum_{i=1}^{K} \underline p_{m,i}\xi_{m,i,k}+\sigma^2\right)\\&-\log_2\left(\sum_{i=1,i\neq k}^{K} \underline p_{m,i}\xi_{m,i,k}+\sigma^2\right),\forall m\in\mathcal M, k\in\mathcal K.  
\end{aligned}
\end{equation}

Besides, to deal with the non-convex terms $\sqrt{\bar t_m}$ and $\sqrt{\underline t_m}$ in constraints (\ref{Problema_pt}) and (\ref{Problemb_pt}), we introduce auxiliary variables $\bar \tau_m$ and $\underline \tau_m$.
Accordingly, we rewrite constraints (\ref{Problema_pt}) and (\ref{Problemb_pt}) as (\ref{upd_aux}), (\ref{downd_aux}), and the following constraints: 
\vspace*{-0.5\baselineskip} 
\begin{equation}
\label{e_up}
    \bar \tau_m\geq\sqrt{\bar t_m},
\end{equation}
\begin{equation}
\label{e_down}
    \underline \tau_m\geq\sqrt{\underline t_m},
\end{equation}
\begin{equation}
\label{hh1}
    \log_2 e~ Q^{-1}\left(\varepsilon_{\mathrm{th}}\right)+\sum_{m=1}^{M}\alpha_{m,k}\left(\frac{\bar{\lambda}_k}{\sqrt{ B_0}\bar \tau_m}-{\sqrt{ B_0}\bar \tau_m}\overline d_{m,k}\right)\leq 0 , 
\end{equation}
\begin{equation}
\label{hh2}
    \log_2 e~ Q^{-1}\left(\varepsilon_{\mathrm{th}}\right)+\sum_{m=1}^{M}\alpha_{m,k}\left(\frac{\underline{\lambda}_k}{\sqrt{ B_0}\underline \tau_m}-{\sqrt{ B_0}\underline \tau_m}\underline d_{m,k}\right)\leq 0.
\end{equation}
To decouple the product of $\overline d_{m, k}$ and $\bar \tau_m$, we have $\overline d_{m, k} \bar \tau_m=\frac{\left(\overline d_{m, k}+\bar \tau_m\right)^2-\left(\overline d_{m, k}-\bar \tau_m\right)^2}{4}$.
Similarly, to decouple the product of $\underline d_{m, k}$ and $\underline \tau_m$, we have $\underline d_{m, k} \underline \tau_m=\frac{\left(\underline d_{m, k}+\underline \tau_m\right)^2-\left(\underline d_{m, k}-\underline \tau_m\right)^2}{4}$.
Accordingly,  constraints (\ref{hh1}) and (\ref{hh2}) are equivalently  expressed as 
\vspace*{-0.5\baselineskip} 
\begin{equation}
\begin{aligned}
&\log_2 e~ Q^{-1}\left(\varepsilon_{\mathrm{th}}\right)+\sum_{m=1}^{M}\alpha_{m,k}\left(\frac{\bar{\lambda}_k}{\sqrt{ B_0}\bar \tau_m}\right.+\frac{\sqrt{ B_0}}{4}\\&\left.\left({\left(\overline d_{m, k}-\bar \tau_m\right)^2 -\left(\overline d_{m, k}+\bar \tau_m\right)^2}\right)\right)\leq 0 , \forall k\in\mathcal K,    
\end{aligned}
\label{aprox11}
\end{equation}
\begin{equation}
\begin{aligned}
&\log_2 e~ Q^{-1}\left(\varepsilon_{\mathrm{th}}\right)+\sum_{m=1}^{M}\alpha_{m,k}\left(\frac{\underline{\lambda}_k}{\sqrt{ B_0}\underline \tau_m}\right.+\frac{\sqrt{ B_0}}{4}\\&\left.\left({\left(\underline d_{m, k}-\underline \tau_m\right)^2-\left(\underline d_{m, k}+\underline \tau_m\right)^2}\right)\right)\leq 0 , \forall k\in\mathcal K.
\end{aligned}
\label{aprox22}
\end{equation}
Next, for control stability constraint (\ref{Problemg}), we re-express it as (\ref{au1}), (\ref{au2}), and (\ref{a_iteration}).
By combining the above transformations, problem (P4) is equivalently re-expressed as
\vspace*{-0.5\baselineskip} 
\begin{align}
\text{(P6)}&:\underset{\substack{\boldsymbol{p}, \boldsymbol{t}, c, \bar{\tau}_m, \underline{\tau}_m, \boldsymbol{\overline{D}}, \boldsymbol{\underline{D}}}}{\text{min}}~ T\nonumber \label{Problem_pt_approximate}\\
\text{s.t.}
&~ \text{(\ref{Problemh}), (\ref{Problemi}), (\ref{Problemc}), (\ref{Problemd}), (\ref{au1}), (\ref{au2}), (\ref{a_iteration})}, \notag\\
&~ \text{(\ref{upd_aux}), (\ref{downd_aux}), (\ref{e_up}), (\ref{e_down}), (\ref{aprox11}), \text{and} (\ref{aprox22}).} \notag
\end{align}

Problem (P6) is non-convex due to the non-convexity of  (\ref{au1}), (\ref{upd_aux}), (\ref{downd_aux}), (\ref{e_up}), (\ref{e_down}), (\ref{aprox11}), and (\ref{aprox22}).
Therefore, we employ the SCA technique to deal with the above constraints in an iterative manner.
In each inner iteration $\underline{o}\geq1$, let $T^{(\underline{o})}$, $\overline d_{m,k}^{(\underline{o})}$, $\underline d_{m,k}^{(\underline{o})}$, $\bar p_{m,k}^{(\underline{o})}$, $\underline p_{m,k}^{(\underline{o})}$, $\bar \tau_{m}^{(\underline{o})}$, and $\underline \tau_{m}^{(\underline{o})}$ denote the local points of $T$,  $\overline d_{m,k}$, $\underline d_{m,k}$, $\bar p_{m,k}$, $\underline p_{m,k}$, $\bar \tau_{m}$, and $\underline \tau_{m}$,  respectively.
Based on the first-order Taylor expansion, the non-convex constraints in (\ref{au1}), (\ref{e_up}), and (\ref{e_down}) are approximated as (\ref{T_approximate_2}), (\ref{e_up_sca}), and (\ref{e_down_sca}) in the following, respectively.
\vspace*{-0.5\baselineskip} 
\begin{equation}
\label{T_approximate_2}
    -T^{(\underline{o})^2}+c+\left(-2 T^{(\underline{o})}\right)\left(T-T^{(\underline{o})}\right) \leq 0,
\end{equation}
\begin{equation}
\label{e_up_sca}
\sqrt{\bar t_m^{(\underline{o})}} + \frac{1}{2} \cdot (\bar t_m - \bar t_m^{(\underline{o})}) \cdot (\bar t_m^{(\underline{o})})^{-1/2} \leq \bar \tau_m, \ \forall m \in \mathcal{M},
\end{equation}
\begin{equation}
\label{e_down_sca}
\sqrt{\underline t_m^{(\underline{o})}} + \frac{1}{2} \cdot (\underline t_m - \underline t_m^{(\underline{o})}) \cdot (\underline t_m^{(\underline{o})})^{-1/2} \leq \underline \tau_m, \ \forall m \in \mathcal{M}.
\end{equation}
Besides, constraints (\ref{upd_aux}), (\ref{downd_aux}),  (\ref{aprox11}), and (\ref{aprox22}) are approximated as (\ref{re_pta}), (\ref{re_ptb}), (\ref{abc}), and (\ref{cba}) at the top of the next page, respectively.

\begin{figure*}[pt]
\normalsize
\setcounter{MYtempeqncnt}{\value{equation}}
\setcounter{equation}{55}

\begin{equation}
\label{re_pta}
\begin{aligned}
    \overline d_{m,k}\!\leq\!\log_2\!\left(\!\sum_{i=1}^{K} \bar p_{m,i}\xi_{m,i,k}+\sigma^2\!\right)\!-\!\log_2\!\left(\!\sum_{\substack{i=1\\i\neq k}}^{K} \bar p_{m,i}^{(\underline{o})}\xi_{m,i,k}+\sigma^2\!\right)\!\sum_{\substack{i=1\\i\neq k}}^{K}\!\left(\bar p_{m,i}\!-\!\bar p_{m,i}^{(\underline{o})}\right)\!\!\left(\!\frac{\log_2 e~\xi_{m,i,k}}{\sum_{\substack{i=1\\i\neq k}}^{K} \bar p_{m,i}^{(\underline{o})}\xi_{m,i,k}+\sigma^2}\!\right),
    \forall m\in\mathcal M, k\in\mathcal K.
\end{aligned}
\end{equation}
\begin{equation}
\label{re_ptb}
\begin{aligned}
    &\underline d_{m,k}\!\leq\!\log_2\!\left(\!\sum_{i=1}^{K} \underline p_{m,i}\zeta_{m,i,k}+\sigma^2\!\right)\!-\!\log_2\!\left(\!\sum_{\substack{i=1\\i\neq k}}^{K} \underline p_{m,i}^{(\underline{o})}\zeta_{m,i,k}+\sigma^2\!\right)\!\sum_{\substack{i=1\\i\neq k}}^{K}\!\left(\underline p_{m,i}\!-\!\underline p_{m,i}^{(\underline{o})}\right)\!\!\left(\!\frac{\log_2 e~\zeta_{m,i,k}}{\sum_{\substack{i=1\\i\neq k}}^{K} \underline p_{m,i}^{(\underline{o})}\zeta_{m,i,k}+\sigma^2}\!\right),\forall m\in\mathcal M, k\in\mathcal K.    
\end{aligned}
\end{equation}  
\begin{equation}
\begin{aligned}
    & \log_2 e~ Q^{-1}\left(\varepsilon_{\mathrm{th}}\right)+\sum_{m=1}^{M}\alpha_{m,k}\left(\frac{\bar{\lambda}_k}{\sqrt{ B_0}\bar \tau_m}+\frac{\sqrt{ B_0}}{4}\left({\left(\overline d_{m, k}-\bar \tau_m\right)^2}\right)\right)
    -\sum_{m=1}^{M}\alpha_{m,k}\frac{\sqrt{B_0}}{4}
    \cdot\left(\left(\overline d_{m, k}^{(\underline{o})}+\bar{e}_{m}^{(\underline{o})}\right)^2
    +2\left(\overline d_{m, k}-\overline d_{m, k}^{(\underline{o})}\right)\right.\\
    & \left.\cdot \left(\overline d_{m, k}^{(\underline{o})}+\bar{e}_{m}\right)+2\left(\bar{e}_{m}-\bar{e}_{m}^{(\underline{o})}\right)\cdot\left(\overline D_{m, k}+\bar{e}_{m}^{(\underline{o})}\right)\right)
    \leq 0 , \forall k\in\mathcal K.    
\end{aligned}
\label{abc}
\end{equation}
\begin{equation}
\begin{aligned}
    &\log_2 e~ Q^{-1}\left(\varepsilon_{\mathrm{th}}\right)+\sum_{m=1}^{M}\alpha_{m,k}\left(\frac{\underline{\lambda}_k}{\sqrt{ B_0}\underline \tau_m}+\frac{\sqrt{ B_0}}{4}\left({\left(\underline d_{m, k}-\underline \tau_m\right)^2}\right)\right)
    -\sum_{m=1}^{M}\alpha_{m,k}\frac{\sqrt{B_0}}{4}
    \cdot\left(\left(\underline d_{m, k}^{(\underline{o})}+\underline{e}_{m}^{(\underline{o})}\right)^2+2\left(\underline d_{m, k}-\underline d_{m, k}^{(\underline{o})}\right)\right.\\
    & \left.\cdot\left(\underline d_{m, k}^{(\underline{o})}+\underline{e}_{m}\right)+2\left(\underline{e}_{m}-\underline{e}_{m}^{(\underline{o})}\right)\cdot\left(\underline D_{m, k}+\underline{e}_{m}^{(\underline{o})}\right)\right)
    \leq 0 , \forall k\in\mathcal K.    
\end{aligned}
\label{cba}
\end{equation}
\hrulefill
\end{figure*}
Accordingly, we approximate problem (P5) as (P7.$\underline{o}$) in the $\underline{o}$-th inner iteration for SCA, which is convex and thus is optimally solvable by convex solvers such as CVX.
\vspace*{-0.5\baselineskip} 
\begin{align}
(\text{P7.}\underline{o})&:\underset{\substack{\boldsymbol{p}, \boldsymbol{t}, c, \bar{\tau}_m, \underline{\tau}_m, \boldsymbol{\overline{D}}, \boldsymbol{\underline{D}}}}{\text{min}}~ T \nonumber  \label{Problem_pt_approximate}\\
\text{s.t.}
&~ \text{(\ref{Problemh}), (\ref{Problemi}), (\ref{Problemc}), (\ref{Problemd}), (\ref{a_iteration}), (\ref{T_approximate_2}), (\ref{e_up_sca}), (\ref{e_down_sca})},  \nonumber\\
&~ \text{(\ref{re_pta}), (\ref{re_ptb}), (\ref{abc}), \text{and} (\ref{cba}).}\notag
\end{align}

Notice that the first-order approximations in the left-hand-side of (\ref{T_approximate_2}), (\ref{re_pta}), (\ref{re_ptb}), (\ref{e_up_sca}), (\ref{e_down_sca}), (\ref{abc}), and (\ref{cba}) serve as upper bounds of those in (\ref{au1}), (\ref{upd_aux}), (\ref{downd_aux}), (\ref{e_up}), (\ref{e_down}), (\ref{aprox11}), and (\ref{aprox22}), respectively. Therefore, solving the series of problems in ($\text{P7.}\underline{o}$) is ensured to result in monotonically decreasing objective values.
Therefore, by solving problem ($\text{P7.}\underline{o}$) in an iterative manner, a converged solution of problem (P5) or equivalently (P6) is obtained.

With problem (P2) and (P5) solved, problem (P1.1) is finally solved by solving them alternately.
In particular, we first fix transmission power $\{\boldsymbol{p}\}$ and solve problem \text{(P2)} to obtain the optimized solution  $\{\boldsymbol{t}, \boldsymbol{\alpha}\}$, and then obtain the updated solution $\{\boldsymbol{t}, \boldsymbol{p}\}$ via solving problem \text{(P5)} under given $\{\boldsymbol{\alpha}\}$.
As the solutions to problems (P2) and (P5) are both converged, the alternating optimization-based algorithm effectively ensures that control latency does not increase. The iterations will terminate when the reduction in the objective value is less than a set threshold. Therefore, the convergence of Algorithm 1 for solving problem (P1.1) is guaranteed.
After problem (P1.1) is solved, we round \(\boldsymbol{\alpha}\) from the relaxed continuous domain \(\alpha_{m,k} \in [0,1]\) back to the binary set \(\alpha_{m,k} \in \{0,1\}\). This step ensures that the original binary constraint (\ref{Problema}) is satisfied in the final solution, following the relaxation introduced in (\ref{alpha_relax}).
After the rounding process, we may need to optimize $\boldsymbol{t}$ and $\boldsymbol{p}$ again. 

\section{Numerical Results}
\begin{figure}[!htbp]
\centering
\includegraphics[width=2.5 in]{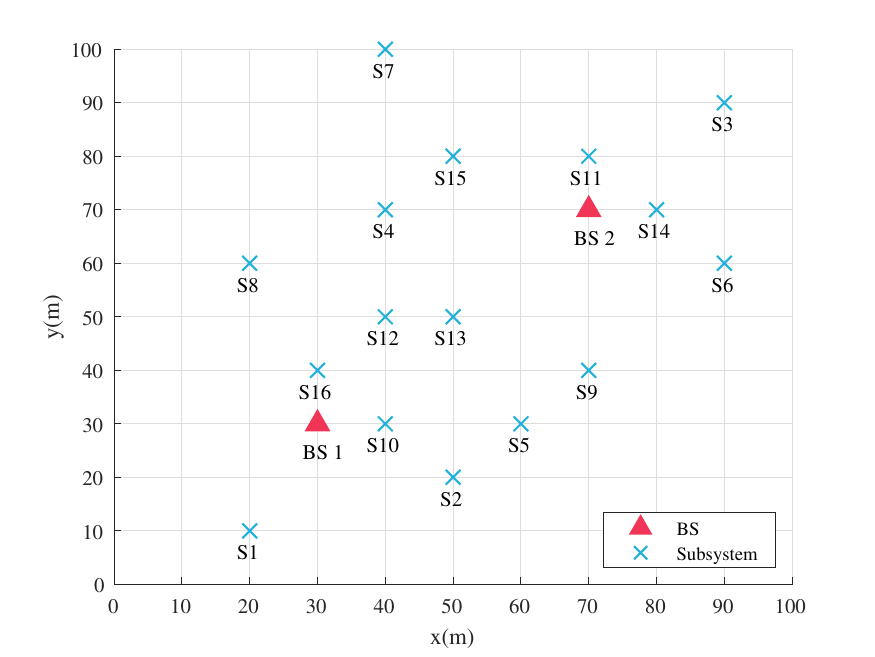}
\caption{The considered wireless networked control system, where 2 BSs and 16 subsystems are placed in an area.}
\label{fig:distribution}
\end{figure}

In this section, we evaluate the performance of our proposed design via numerical simulations. 
Unless otherwise specified, the parameter settings of the simulations are as follows.
We set $M=2$ BSs, $N=32$ antennas, and $K=16$ plants.
The considered network is shown in Fig.~\ref{fig:distribution}, where two BSs are located at (30,30) m and (70,70) m, respectively.
The considered subsystems are located uniformly over [0, 100] m.
For control model, we consider $a=2$ state values with $\mathbf{R}_k=[1,0;0,1]$, $\mathbf{A}_k=[1,1;0,1]$, $\mathbf{B}_k=[1,0;0,1]$, $\mathbf{Q}_k=[1,0;0,1]$, and $\eta_k=0.8$, $\forall k \in \mathcal{K}$\cite{control_model}.
For communication model, we set $\underline{P}_m=5\text{W}, \forall m \in \mathcal{M}$, and $\bar{P}_k=0.5\text{W}, \forall k \in \mathcal{K}$.
Meanwhile, the bandwidth is set as $B_0=10$MHz and the noise power spectral density is set as $N_0=-110$dBm/Hz.
The path loss between BS and plant is modeled as $\beta_0\left(\frac{d_{m,k}}{d_0}\right)^{-\xi}$, where $\beta_0=-30$dB represents the path loss at the reference distance $d_0=1$m, $d_{m,k}$ is the distance between plant $k$ and BS $m$, and $\xi$ is the path loss exponent.
The channel vector $\boldsymbol{g}_{m,k}$ is randomly generated from  Rayleigh fading with the average channel powers set according to the path loss.
For computation process, we set $S_1=S_2=1000$cycles/bit and  $F_{m}=\num{1e9}$cycles/s.
We assume that $\bar{\lambda}_k=\hat{\lambda}_k=\underline{\lambda}_k=500$bits, $\forall k \in \mathcal{K}$.
\vspace*{-0.5\baselineskip} 
\subsection{Benchmark Schemes for Comparison}
In this subsection, we introduce the following four benchmark schemes for performance comparision:

\begin{itemize}
    \item Association optimization only: 
    We only optimize BS-sensor/actuator association with equal transmission power allocation.
    This corresponds to solving problem (P2) with equal transmission power $\bar{p}_{m, k}$ and $\underline{p}_{m, k}$ allocated to each plant as $\bar{p}_{m, k}=\bar{P}_k=0.5\text{W}$ and $\underline{p}_{m, k}=\frac{\underline{P}_m}{\sum_{k \in \mathcal{K}} \alpha_{m, k}}, \forall m \in \mathcal{M}, \forall k \in \mathcal{K}$.
    \item Resource allocation only:
    We only optimize the transmission power allocation between plants and BSs, with predetermined BS-sensor/actuator association.
    In particular, the association between BSs and sensors/actuators is determined based on proximity, such that each subsystem is associated with the nearest BS. 
    Accordingly, the transmission power allocation is obtained by solving problem (P5). 

\item FDMA design:
The FDMA protocol is used for communication among BSs, with the bandwidth $B_0$ uniformly allocated to subsystems at each BS to ensure orthogonal frequency resources. Each sample period $T$ is divided into three slots for uploading ($\bar t$), computing ($\hat t$), and downloading ($\underline t$).
In the uploading slot, subsystems transmit data to BSs, with $\bar t=\mathrm{max}{\bar t_m}, \forall m\in \mathcal{M}$. Following this, BSs process the data during the computation slot, where $\hat t=\mathrm{max}{\hat t_m}, \forall m\in \mathcal{M}$. Finally, BSs send commands back in the downloading slot, with $\underline t=\mathrm{max}{\underline t_m}, \forall m\in \mathcal{M}$.

The closed-loop control latency $T=\bar t+\hat t+\underline t$ is minimized under communication constraints (\ref{Probleme})-(\ref{Problemf}), control stability (\ref{Problemf}), association constraints (\ref{Problemi})-(\ref{Problemb}), and computation constraints $F_{m}\hat{t}m\ge \sum{k \in \mathcal{K}}\alpha_{m,k}S_{k}\hat{\lambda}_{k}, \forall m\in\mathcal{M}$. Decision variables include transmission power, BS-sensor/actuator association, and time allocation, solved similarly to (P1).
\end{itemize}
\vspace*{-0.5\baselineskip}

\subsection{Simulation Results}
\begin{figure}[!htbp]
\centering
\includegraphics[width=2.5 in]{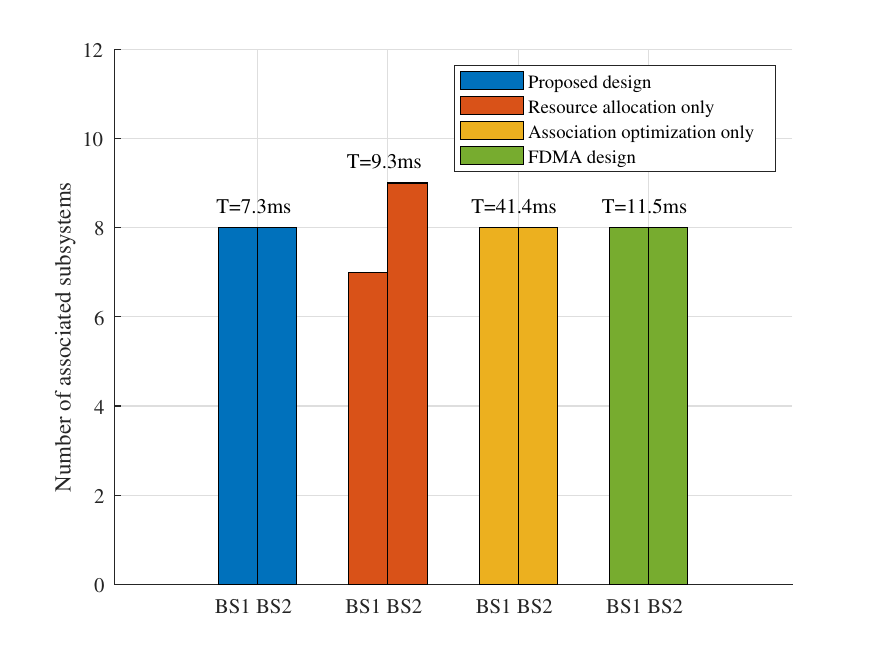}
\caption{Number of subsystems associated with each BS under equal transmission power and computation frequency at BSs.\textsuperscript{3}}
\label{fig:equal}
\end{figure}

Fig.~\ref{fig:equal} shows the number of subsystems associated with each BS under our proposed design and the three benchmark schemes.
The transmission power of BSs are set as $\underline P_1=\underline P_2=5$W and the computation frequency of BSs are set as $F_1=F_2=\num{1e9}$cycles/s.
It is observed that our proposed design achieves lower closed-loop control latency than the other benchmark schemes via efficient user association and resource allocation.
Furthermore, it is observed that less subsystems are associated with BS1 under resource allocation only scheme.
This is due to the fact that for resource allocation only scheme, the subsystems are simply associated with their nearest BS, which leads to disproportionately heavy computation and communication load for BS2.
\footnotetext[3]{S1, S2, S5, S8, S10, S12, S13, and S16 are associated with BS1 and the other subsystems are associated with BS2 in our proposed design, the resource allocation only scheme and the FDMA design. S1, S2, S5, S8, S10, S12, and S16 are associated with BS1 and the other subsystems are associated with BS2 in the resource allocation only scheme.}    

\begin{figure}[!htbp]
\centering
\includegraphics[width=2.5 in]{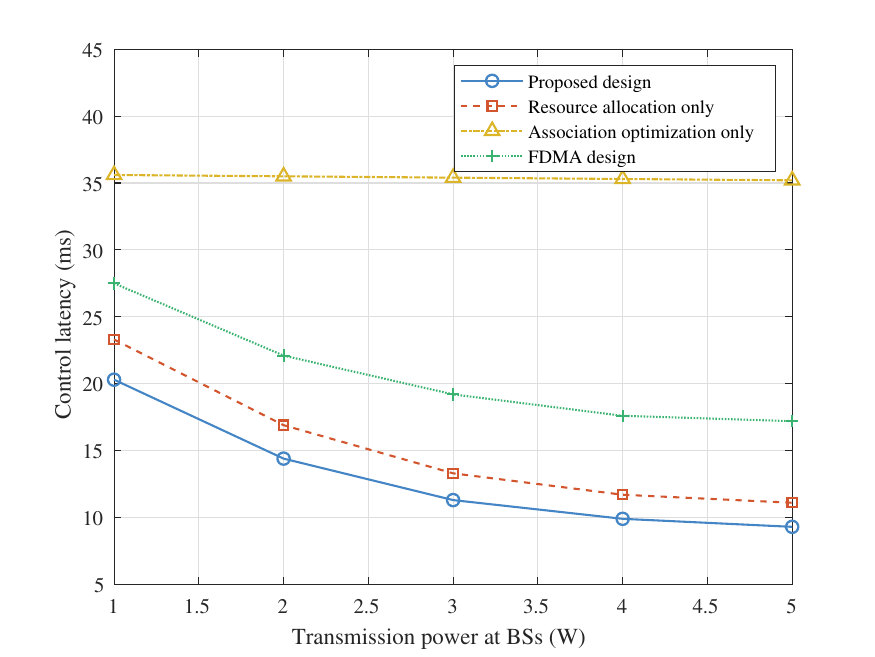}
\caption{Closed-loop control latency versus transmission power at BSs.}
\label{fig:transmission_power_latency}
\end{figure}
\vspace*{-0.5\baselineskip} 

Fig.~\ref{fig:transmission_power_latency} shows the closed-loop control latency versus the transmission power $\underline P_m$ with our proposed algorithm and three benchmarks.
It is observed that the control latency decreases with the improvement of transmission power $\underline P_m$ of BS in our proposed design, the FDMA design, and the resource allocation only scheme. 
Conversely, the association optimization only scheme is observed to have minimal impact on the control latency, which remains nearly static regardless of changes in the BS power levels. 
This indicates that without specific adjustments to power distribution, merely altering association or evenly distributing resources does not effectively reduce the latency in this network.

\begin{figure}[!htbp]
\centering
\includegraphics[width=2.5 in]{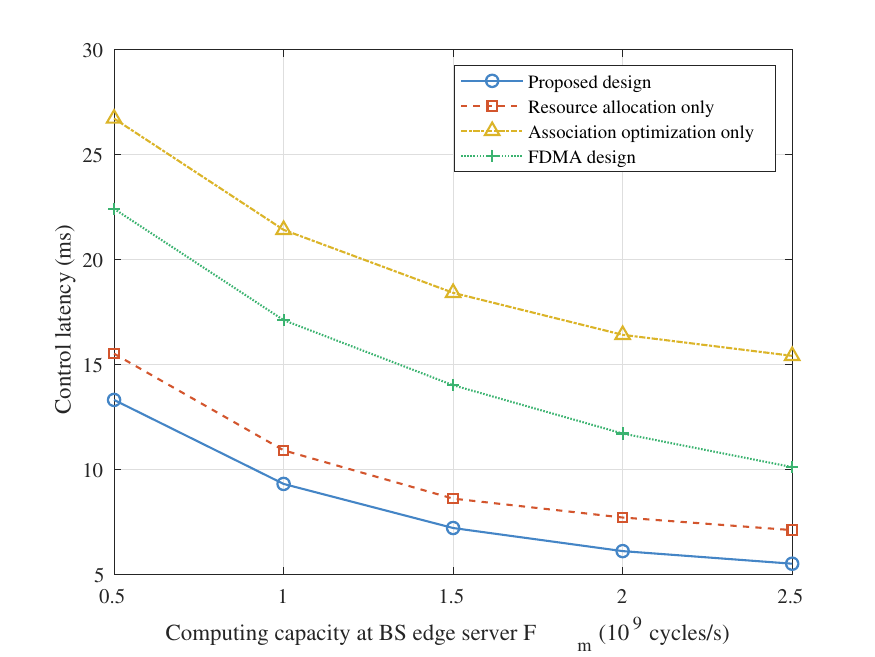}
\caption{Closed-loop control latency versus computing capability of edge server $F_m$.  }
\label{fig:computing_frenquency}
\end{figure}

\begin{figure}[!htbp]
\centering
\includegraphics[width=2.5 in]{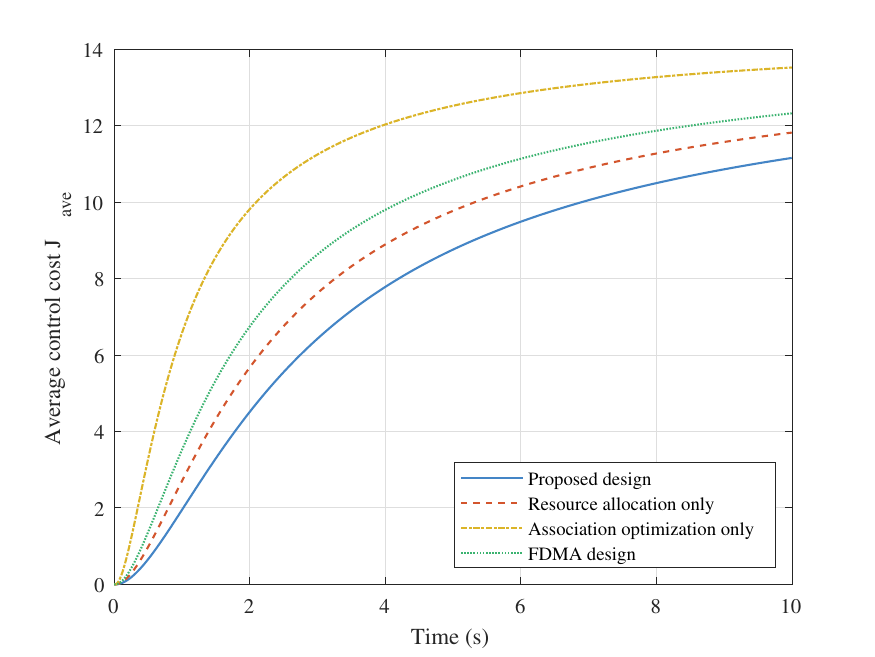}
\caption{Average control cost versus time.  }
\label{fig:control_cost}
\end{figure}

Fig.~\ref{fig:computing_frenquency} shows the closed-loop control latency versus the computation frequency of edge server $F_m$ achieved by different designs.
It is observed that with $F_m$ increasing, the closed-loop control latency decreases for all schemes.
This is due to the fact that higher computation frequencies enhance the processing capabilities of the network, enabling quicker data processing.
Next, it is observed that when the computing capabilities are limited, our proposed design outperforms the FDMA design. 
This means that our design has a significant advantage in practical scenarios where the computing resources are constrained, effectively managing to maintain lower latency levels under such conditions.

Finally, Fig.~\ref{fig:control_cost} shows the control performance of this system in terms of average control cost $J_{a v e}=\sum_{k=1}^K\frac{\sum_{n=1}^N\|\mathbf{x}_k[n]\|^2}{N}$\cite{36}\cite{37}.
It is observed that initially, $J_{ {ave }}$ increases across all schemes, indicating the system's transition toward a stable state.  
As time progresses, both the system state and average control costs tend to stabilize.
Notably, the proposed design consistently outperforms other benchmarks in terms of lower control costs, underscoring the efficacy of our proposed design.
The proposed design's advantage suggests that jointly optimizing the resource utilization and BS-sensor/actuator association is critical in achieving efficient wireless network control performance over time.

\section{Conclusion}
In this paper, we investigated a wireless networked control system with multiple BSs managing subsystems of plants, sensors, and actuators. Sensors offload data to BSs for MEC processing, and BSs send control signals back using TDMA and massive MIMO for efficient communication and computation. The study formulated a closed-loop control latency minimization problem, optimizing BS-sensor/actuator associations and resource allocation under HRLLC constraints. To solve the non-convex problem, an alternating optimization algorithm with SCA was proposed. Numerical results shown that the proposed design achieves lower latency and better stability than benchmark schemes.
\bibliographystyle{IEEEtran}
\bibliography{ref1}

\end{document}